\documentclass{article}

\usepackage{graphicx}
\usepackage{amsthm}
\usepackage{amsmath}
\usepackage{amssymb}
\usepackage{epsfig}
\usepackage{enumerate}
\usepackage{algorithm}
\usepackage[noend]{algpseudocode}
\usepackage{multirow}
\usepackage{authblk}

\theoremstyle{definition}
\newtheorem{example}{Example}[section]

\begin{document}

\title{An Improved One-to-All Broadcasting in Higher Dimensional Eisenstein-Jacobi Networks}
\author{Zaid Hussain\\ zhussain@cs.ku.edu.kw}
\affil{Computer Science Department\\ College of Computing Sciences and Engineering\\ Kuwait University}

\maketitle

\begin{abstract}
Recently, a higher dimensional Eisenstein-Jacobi (EJ) networks, $EJ_\alpha^{(n)}$, has been proposed in \cite{Hussain2016}, which is shown that they have better average distance with more number of nodes than a single dimensional EJ networks. Some communication algorithms such as one-to-all and all-to-all communications are well known and used in interconnection networks. In one-to-all communication, a source node sends a message to every other node in the network. Whereas, in all-to-all communication, every node is considered as a source node and sends its message to every other node in the network. In this paper, an improved one-to-all communication algorithm in $EJ_\alpha^{(n)}$ networks is presented. The paper shows that the proposed algorithm achieves a lower average number of steps to receiving the broadcasted message. In addition,  since the links are assumed to be half-duplex, the all-to-all broadcasting algorithm is divided into three phases. The simulation results are discussed and showed that the improved one-to-all algorithm achieves better traffic performance than the well-known one-to-all algorithm and has 2.7\% less total number of senders.
\end{abstract}


\section{Introduction}
\label{sectionIntroduction}
Multiprocessors are categorized into two types. The first type is called distributed memory multiprocessors system where the communications between the processors are performed over an interconnection network since every processor has its own memory unit. The other type is called shared-memory multiprocessors system where the processors communicate through a common memory unit. Thus, the topology of interconnection networks plays critical roles in in achieving high performance in distributed memory multiprocessors. There are many popular interconnection networks such as  Hypercubes \cite{hayes1989hypercube,hayes1986architecture}, Generalized Hypercubes \cite{bhuyan1984generalized}, Twisted Cubes \cite{hilbers1987twisted}, Cube Connected Cycle \cite{preparata1981cube}, $k$--ary $n$--cube \cite{bose1995lee}, and Torus \cite{dally1986torus}.

One of the efficient interconnection networks called Eisentein-Jacobi networks, simply EJ, introduced in \cite{flahive2010topology,martinez2008modeling}. EJ networks are symmetric and 6-regular networks where each node in the network is connected to 6 neighbors. They are based on EJ integers, which are used to implement error-free of radix-3 Fast Fourier Transform (FFT) algorithms and efficient algorithms for complex multiplication \cite{dimitrov1999eisenstein}. EJ networks are known to be a generalization of hexagonal networks developed in \cite{chen1990addressing,dolter1991performance,kandlur1991reliable}. Thus, the applications on hexagonal networks can also be applied on EJ networks. As an extension, in \cite{Hussain2016} a higher dimensional EJ network has been developed based on the cross products between the lower dimensional EJ networks. the next section describes the formal definition of this network.

The design of efficient communication algorithms for parallel computing systems has been a hot topic. For instance, the problems related to image processing and computer vision \cite{arabnia1990parallel,arabnia1995distributed,arabnia1987arbitrary,arabnia1987transputer,arabnia1989transputer,bhandarkar1995hough,wani2003parallel} are solved based on the implementation of parallel algorithms. Furthermore, some studies on one-to-all and all-to-all broadcasting can be found in\cite{duato2003interconnection,hussain2015edge,touzene2015all}. In one-to-all broadcasting, a source node sends its message to every other node in the network. Whereas, in all-to-all broadcasting, all nodes in the network send their messages to every other node in the network. Some applications use these types of communications such as matrix transpose, matrix multiplication, some parallel database join operations, etc.

In \cite{Hussain2016}, the implementation of one-to-all broadcasting was semi-parallelized. That is, each node forwards the received message to its neighbors on the same dimension where the message was received. In this paper, the broadcasting algorithms are studied and enhanced to be fully parallelized to achieve better traffic distributions in higher dimensional EJ network.

The structure of the paper is as follows. In Section \ref{sectionBackground}, the topological properties of EJ network are briefly described. The previous work related to this paper is discussed in Section \ref{sectionPrev}. Section \ref{sectionBroadcast} presents an improved one-to-all broadcasting algorithm and used to implement the all-to-all broadcasting algorithm. In Section \ref{sectionAnalysis}, a comparative analysis between the previous and the proposed algorithms is illustrated. Simulation results are illustrated and discussed in Section \ref{sectionSimulation}. Finally,
the paper is concluded in Section \ref{sectionConclusion}.

\section{Background}
\label{sectionBackground}
In this section, the important topological properties of Eisenstein-Jacobi networks are reviewed in subsection \ref{EJnetwork}. Further, the design and the definition of higher dimensional Eisenstein-Jacobi networks are summarized in subsection \ref{higherEJnetworks}.

\subsection{Eisenstein-Jacobi Network ($EJ_\alpha$)\label{EJnetwork}}
Based on the definitions mentioned in \cite{flahive2010topology,huber1994codes}, an EJ integer $\mathbb{Z}[\rho]$ is the subset of complex numbers with real and imaginary parts denoted as
$\mathbb{Z}[\rho] = \{x+y\rho \mid x,y \in \mathbb{Z}\}$ where $\rho = (1+i\sqrt{3})/2$, $i = \sqrt{-1}$, $\rho^2 = -1+\rho$, and $\mathbb{Z} = \{0, 1, 2, \dots\}$. $EJ_\alpha$ networks belong to a family of planer graphs, which can be represented as graph $EJ_\alpha(V,E)$ generated by $\alpha = a+b\rho$ such that $0 \leq a \leq b$ where $V = \mathbb{Z}[\rho]_\alpha$ is the node set and $E = \{(A,B) \in V \times V \mid (A-B) \equiv \pm 1, \pm \rho, \pm \rho^2 \ mod \ \alpha \}$ is the link set represents the connections between the nodes.

An $EJ_\alpha$ network generated by $\alpha = a+b\rho \neq 0$ is based on quotient rings of EJ integers and it contains a total number of nodes equal to $N(\alpha) = a^2 + b^2 + ab$, called norm, which is the number of the elements in the residue class modulo $\alpha$ \cite{huber1994codes}. $EJ_\alpha$ networks are 6-regular symmetric networks, i.e., each node in the network has six neighbors. Each node in the network is labeled as $x+y\rho$, which represents the location of EJ integer on the grid. Two nodes $A$ and $B$ are neighbors if $(A-B) \ mod \ \alpha$ is $\pm 1$, $\pm \rho$, or $\pm \rho^2$. The diameter of the network is known as the  shortest distance between two most farthest nodes in the network $EJ_\alpha$.

The distance between any two nodes $A$ and $B$ in the network is defined as:
\begin{eqnarray}
\nonumber D_\alpha(A,B) &=& \min \{|x| + |y| + |z| \ | \ (A - B) \\
& & \equiv x + y\rho + z\rho^2 \ (mod \ \alpha) \}
\end{eqnarray}

Since \begin{math}EJ_\alpha\end{math} is node-symmetric the weight of node $A$, which is the distance of this node from node 0, is defined as:
\begin{eqnarray}
\nonumber W_\alpha(A) &=& \min \{|x| + |y| + |z| \ | \ (A) \\
& & \equiv x + y\rho + z\rho^2 \ (mod \ \alpha) \}
\end{eqnarray}

Given a distance $s$, where $s = 0, 1, 2, \dots, M$ and $M$ is the diameter of the network defined below, the number of nodes at distance $s$ in $EJ_\alpha$ is denoted as $W_{EJ}(s)$ \cite{flahive2010topology}. The following describes the distance distribution of the network.
\begin{eqnarray}
W_{EJ}(s) = \left\{
{\begin{array}{*{20}{l}}
1&{if \ s = 0}\\
{6s}&{if \ 1 \le s < T}\\
{18(M - s)}&{if \ T < s < M}\\
2&{if \ b \equiv a \ (mod \ 3) }\\
 & {and \ s = M}\\
0&{if \ s > M}\\
{N(\alpha ) - R}&{if \ s = T}
\end{array}} \right.
\end{eqnarray}
where $T = (a + b)/2$, $M = (a + 2b)/3$, $R = \sum_{s = 0,s \ne T}^M W_{EJ}(s)$ and the diameter of the network is at most $M$. The value of $W_{EJ}(T)$ depends on whether $T$,$M$ are integers; that is, it depends on the value of $a - b \ (mod \ 6)$. The value of $W_{EJ}(T)$ can be found by subtracting the sum of the weights already listed from the total number of nodes $a^2 + b ^2 + ab$.

There are two types of links in $EJ_\alpha$ networks: regular and wraparound links. The links that connect two nodes within the network grid are called regular links, whereas, the wraparound links connect the nodes located within the network grid with the other node located out of the boundary of the network grid. In other way, the wraparound links can be understood by using $mod \ \alpha$ operation after adding $\pm 1$, $\pm \rho$, or $\pm \rho^2$ to the nodes located at the boundary of the grid to get their neighbors located out of the boundary of the grid. In addition, the wraparound links can be easily seen by placing the $EJ_\alpha$ network at the origin of a grid and consider it as a basic $EJ_\alpha$ network with its center node is 0. Then, making tiles by copying the basic $EJ_\alpha$ network and placing its copies around it.

\begin{figure}[H]
\centering
\includegraphics[scale=0.9]{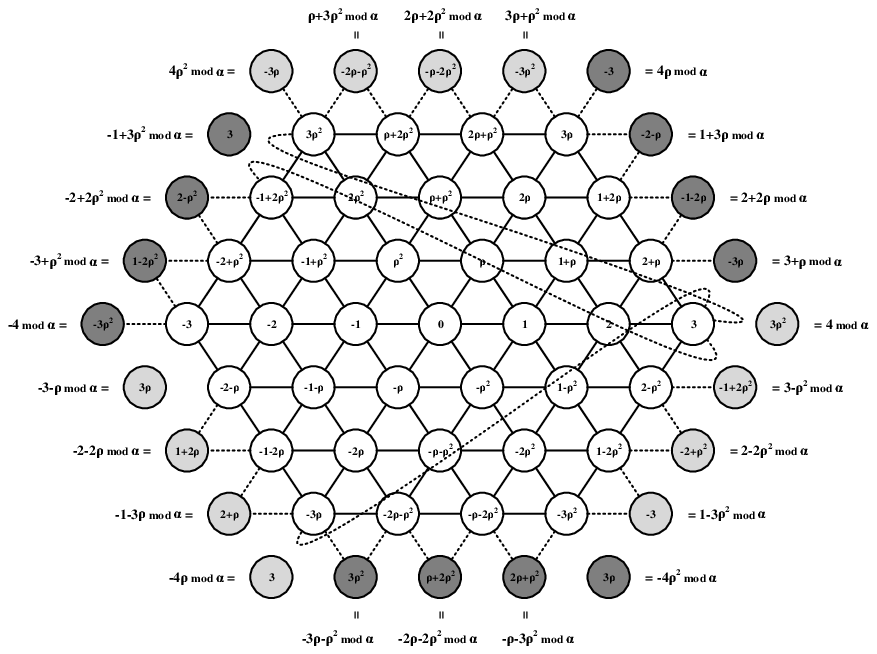}
\caption{$EJ_{3+4\rho}$ with dotted lines as a wraparound links.}
\label{wraparound}
\end{figure}

Figure \ref{wraparound} illustrates the $EJ_\alpha$ network generated by $\alpha = 3+4\rho$ where the solid lines represent the regular links and the dotted lines represent the wraparound links. Note that, in Example \ref{exWraparound}, we have removed the straight dotted lines from node 3 to describe them as wrapped links. Also, we have kept the boundary nodes of the tiles to represent the nodes located out of the boundary of the grid and the rest nodes of the tiles are removed. The nodes in different tiles of the network are represented in different gray colors.

\begin{example}
Consider the node 3 in Figure \ref{wraparound}. The node 3 is connected to node $3+\rho$ through $+\rho$ link, which its corresponding node within the basic grid is $-3\rho$. That is, $3+\rho \ mod \ \alpha = -3\rho$. Similarly, the $+1$ and $-\rho^2$ links of node 3 connect the node 3 to nodes 4 and $3-\rho^2$, respectively. Note that, in respective order, their corresponding nodes in the basic grid are $3\rho^2$ and $-1+2\rho^2$.
\label{exWraparound}
\end{example}

\begin{figure}[!ht]
\centering
\includegraphics[scale=0.35]{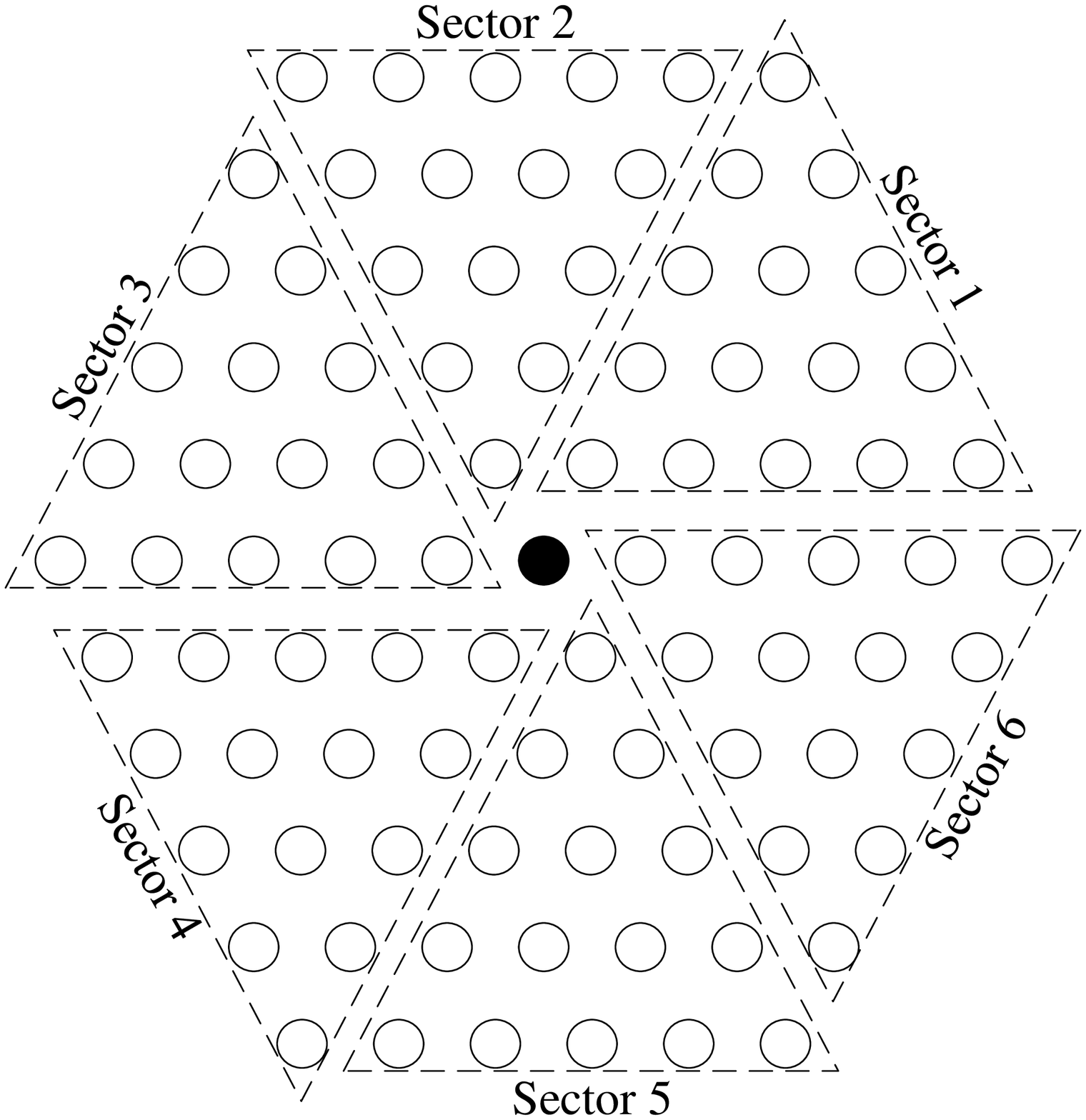}
\caption{The six sectors in $EJ_{5+6\rho}$.}
\label{EJ5+6rho_sectors}
\end{figure}

The network can be divided into six sectors. This division is useful in both one-to-all and all-to-all broadcasting. Figure \ref{EJ5+6rho_sectors} shows an example of the six sectors in $EJ_{5+6\rho}$ network where the black node is the center node of the network, usually node 0. A node $A = x\rho^{j-1} + y\rho^j$ is in sector $j$ for $j$ = 1, 2, 3, 4, 5, and 6.

\subsection{Higher Dimensional EJ Network $EJ^{(n)}_\alpha$\label{higherEJnetworks}}
The higher dimensional $EJ_\alpha^{(n)}$ network was proposed in \cite{Hussain2016}. It is shown that the higher dimensional $EJ_\alpha^{(n)}$ network is formed based on the cross product between the lower dimensional EJ networks as follows:
\begin{eqnarray}
\nonumber EJ_\alpha^{(n)} &=& EJ_\alpha \otimes EJ_\alpha^{(n-1)} \\
& & = EJ_\alpha \otimes \overbrace{(EJ_\alpha \otimes \dots \otimes EJ_\alpha)}^{n-1 \ times}
\end{eqnarray}
where $0 \neq \alpha = a+b\rho \in \mathbb{Z}[\rho]$ and $n$ is the number of dimensions of $EJ_\alpha^{(n)}$.

The cross product between two graphs is explained in \cite{deo2016graph} as follows. Let $G(V,E)$ be the resultant graph from the cross product between two graphs $G_1 = ({V_1},{E_1})$ and ${G_2} = ({V_2},{E_2})$. Then, $G(V,E)$ can be written as ${G_1} \times {G_2}$ where $V = \{ (u,v) \ | \ u \in {V_1},v \in {V_2}\}$ and $E = \{ (( {{u_1},{v_1}} ),( {{u_2},{v_2}} ))|(({u_1},{u_2}) \in {E_1}$ and ${v_1} = {v_2})$ or $(({v_1},{v_2}) \in {E_2}$ and ${u_1} = {u_2})\}$.

\begin{figure}[!ht]
\centering
\includegraphics[scale=0.40]{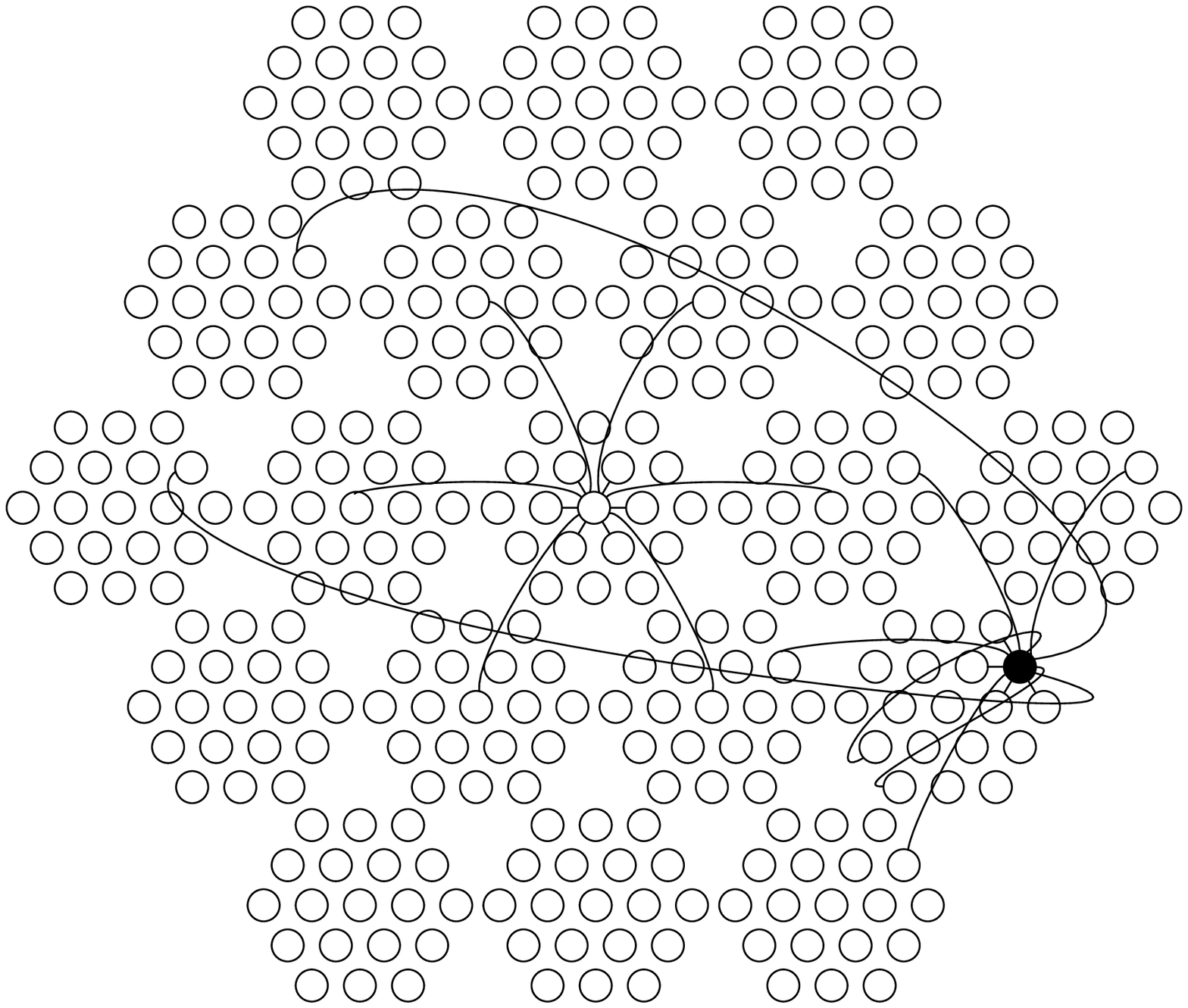}
\caption{$EJ_{2+3\rho}^{(2)}$ with nodes 0 and ($1-\rho^2$,$1+\rho$) and their neighbors. The black node is ($1-\rho^2$,$1+\rho$).}
\label{2+3i_2D_ex}
\end{figure}

The total number of nodes in $EJ_\alpha^{(n)}$ is $N(\alpha)^n$, which is the total number of nodes in a single dimensional EJ network power of $n$. A node in $EJ_\alpha^{(n)}$ is denoted as a set of $n$-tuples with coordinates in $EJ_\alpha$. That is, a node ($x_n + y_n\rho$, $x_{n-1} + y_{n-1}\rho$, \dots, $x_1 + y_1\rho$) is located in the positions $x_n + y_n\rho$ on the first layer (highest or $n^{th}$-dimension) of $EJ_\alpha^{(n)}$, $x_{n-1} + y_{n-1}\rho$ on the second layer of $EJ_\alpha^{(n)}$, and so on until $x_1 + y_1\rho$ on the last layer (lowest or $1^{st}$-dimension) of $EJ_\alpha^{(n)}$. The degree of each node is $6n$. $EJ_\alpha^{(n)}$ can be drawn by placing a copy of $EJ_\alpha^{(n-1)}$ on each node of $EJ_\alpha$. For example, Figure \ref{2+3i_2D_ex} illustrates $EJ_{2+3}^{(2)}$ where the node ($1-\rho^2$, $1+\rho$) is filled with a black color with all of its edges connected to its neighbors and the neighbors of node (0,0) are obvious.

\section{Previous Work}
\label{sectionPrev}
In this section, the known one-to-all broadcasting algorithm is reviewed. After that, the one-to-all broadcasting algorithm for $EJ_\alpha^{(n)}$ is briefly described.

Before describing the algorithm, except the node $0+0i$ (simply 0), there are two types of receiving nodes, the axis nodes $x\rho^p$, for $p=0, 1, 2$ and $x$ is either positive or negative integer; and the non-axis nodes, i.e., the rest of the nodes. The one-to-all broadcasting \cite{albader2012efficient,kandlur1991reliable} in $EJ_\alpha$ can be performed in $M$ steps based on the spanning tree and the six sectors of $EJ_\alpha$ as described in Figure \ref{EJ5+6rho_sectors}, where the numbered links illustrates the steps of the broadcasting. 

Consider the network $EJ_\alpha$. The spanning tree of the network is constructed during the broadcasting process as follows. In the first step, the root node of the spanning tree, usually node 0, sends its message to the six sectors through all neighbor nodes ($1$, $\rho$, $\rho^2$, $-1$, $-\rho$, $-\rho^2$), where each neighbor node is responsible to distribute the message to its sector. From steps 2 to $M$ the algorithm works as follows. In each step, each node receives the message from its parent node will forward the message to its children within its sector. That is, the axis nodes forward the message to two neighbor nodes and the non-axis nodes forward the message to one neighbor node, all located within the same sector.
 
\begin{figure}[H]
\centering
\includegraphics[scale=0.4]{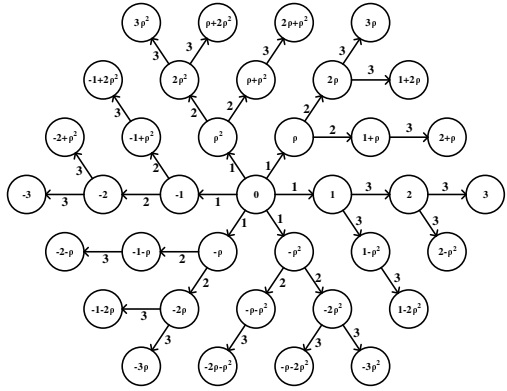}
\caption{One-to-all broadcast in $EJ_{3+4\rho}$.}
\label{3+4rho_bc_1D}
\end{figure}

 For example, consider the network $EJ_{3+4\rho}$. Based on the spanning tree illustrated in Figure \ref{3+4rho_bc_1D},
in the first step, the node 0 sends a message to nodes $1$, $\rho$, $\rho^2$, $-1$, $-\rho$, and $-\rho^2$ through the links numbered 1. The rest of the broadcasting process is explained for sector 6 since all sectors perform the same steps of the algorithm. In the second step, the axis node 1 receives its message from node 0 and then forwards the message to the axis node 2 and the non-axis node $1-\rho^2$ via the links numbered 2. Finally, in the third and last step, the axis node 2 forwards its received message to its neighbor nodes 3 and $2-\rho^2$ through the links numbered 3, whereas, the non-axis node $1-\rho^2$ forwards the received message to its neighbor node $1-2\rho^2$ via the links numbered 3.

\begin{figure}[!ht]
\centering
\includegraphics[scale=0.4]{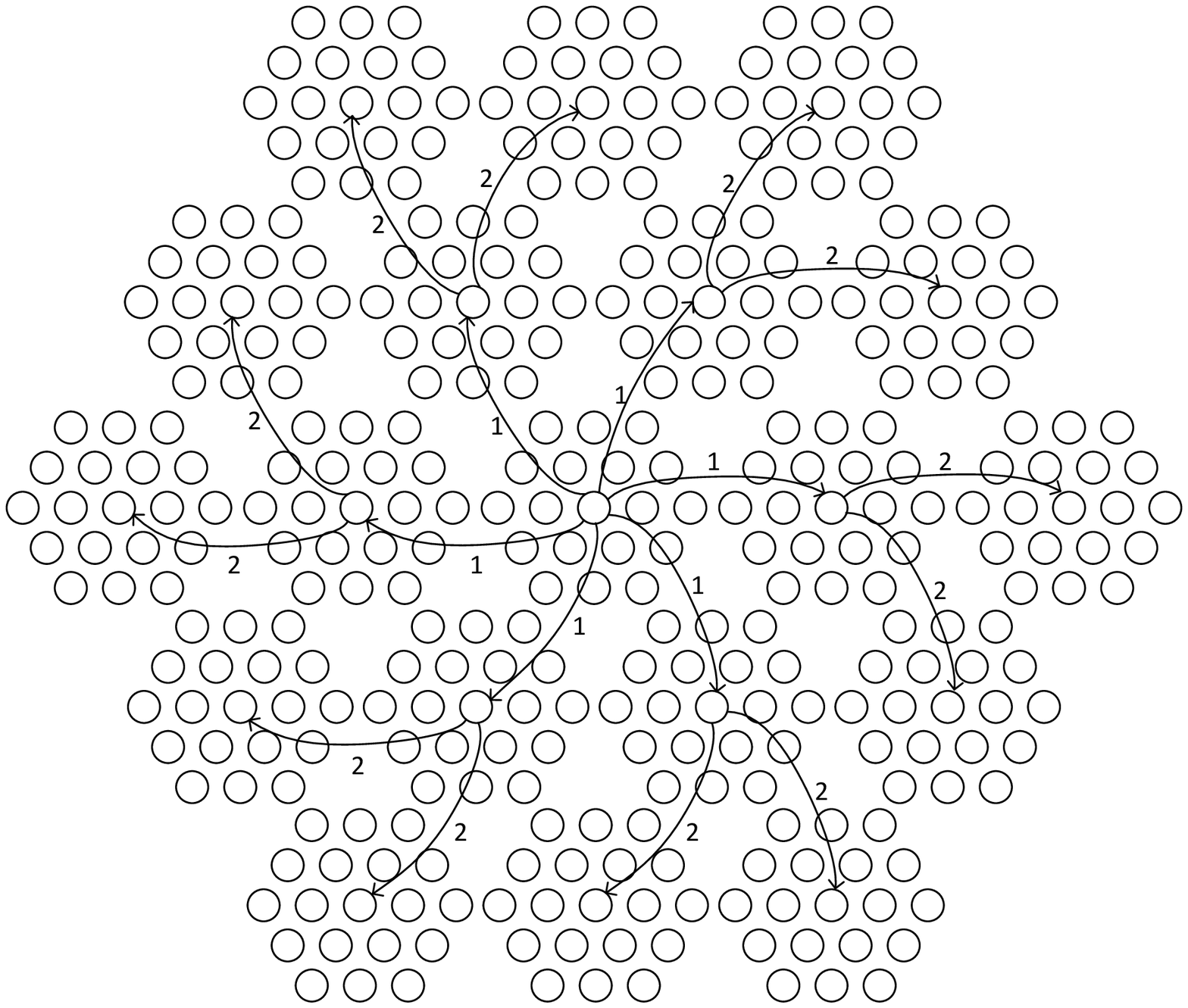}
\caption{First round of one-to-all broadcast in $EJ_{2+3\rho}^{(2)}$.}
\label{2+3rho_bc_2D_round1}
\end{figure}

\begin{figure}[!ht]
\centering
\includegraphics[scale=0.4]{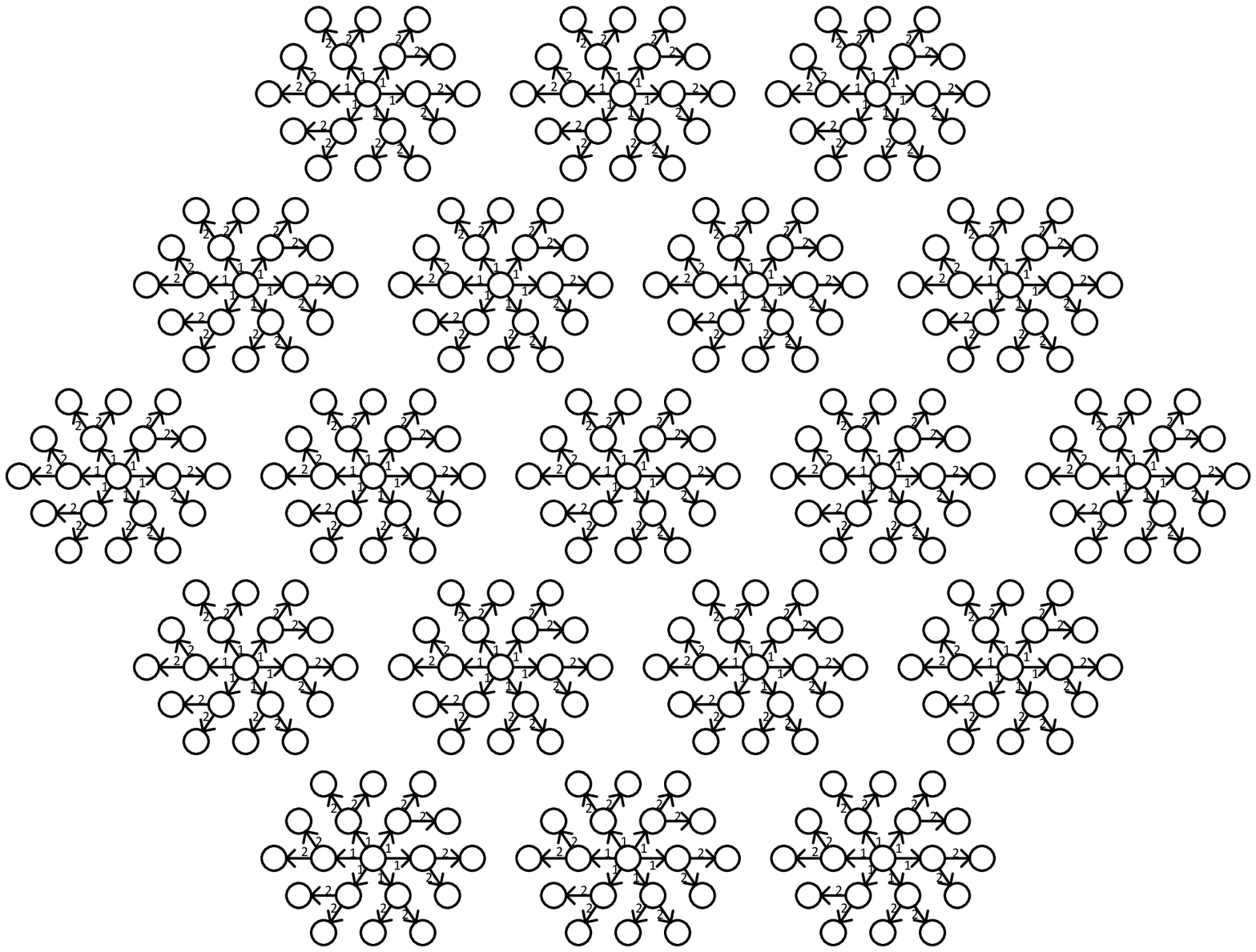}
\caption{Second round of one-to-all broadcast in $EJ_{2+3\rho}^{(2)}$.}
\label{2+3rho_bc_2D_round2}
\end{figure}

In \cite{Hussain2016}, an iterative one-to-all broadcasting algorithm has been developed for $EJ_\alpha^{(n)}$ based on the above described one-to-all broadcasting algorithm for $EJ_\alpha$. The one-to-all broadcasting in $EJ_\alpha^{(n)}$ is divided into $n$ rounds such that each round has $M$ steps. For round 1, the $M$-steps of one-to-all are applied on the $n^{th}$ dimension (first layer). When the round 1 is ended, each center node in $EJ_\alpha^{(n-1)}$ has received the message. Then, the one-to-all is applied on the $(n-1)^{th}$ dimension in round 2. That is, each of $EJ_\alpha^{(n-1)}$ applies the one-to-all broadcast. Repeating this process for $n$ rounds from the highest layer to the lowest layer makes all nodes receive the message. That is, for round $i+1$, where $i = 0, 1, \dots, n-1$, the $M$-steps of one-to-all are applied on every $EJ_\alpha^{(n-i)}$ in the $(n-i)^{th}$ dimension. Figures \ref{2+3rho_bc_2D_round1} and \ref{2+3rho_bc_2D_round2}, in respective order, illustrate the first and second rounds of the one-to-all broadcasting process in $EJ_{2+3\rho}^{(2)}$.


\section{An Improved Broadcasting in $EJ^{(n)}_\alpha$}
\label{sectionBroadcast}
This section proposes the improved one-to-all broadcasting algorithm. Further, the proposed one-to-all algorithm is used in the implementation of all-to-all broadcasting algorithm. For simplicity, the algorithms below are described for the $EJ_\alpha^{(n)}$ networks where $b = a+1$. The same algorithms with little modifications could be used to perform the broadcasting on $EJ_\alpha^{(n)}$ such that $0 \leq a \leq b$.

\subsection{One-to-All Broadcast}
\label{sectionO2A}
The previous and the proposed one-to-all algorithms have the same number of rounds and steps, which is $nM$ where $M$ is the diameter of the network and $n$ is the number of dimensions. The advantage of the proposed one-to-all algorithms over the previous one is that the proposed algorithms distribute the broadcast traffic over the broadcast steps on a relatively more balanced manner. This results in a lower average number of steps needed to receive the broadcasted message.

Algorithms \ref{one2all} and \ref{one2allSector} describe the proposed one-to-all broadcasting algorithm on $EJ_{a+b\rho}^{(n)}$ where $b = a+1$ and $n > 0$. Simply, the node 0 sends its message to all of its neighboring nodes in all dimensions. Then, as in the previous one-to-all algorithm, each receiving node sends the received message to the neighboring nodes in its sector. In addition, the receiving nodes send their message to all of their neighboring nodes in all dimensions that are lower than the receiving dimension.

\begin{figure}[!ht]
\centering
\includegraphics[scale=0.45]{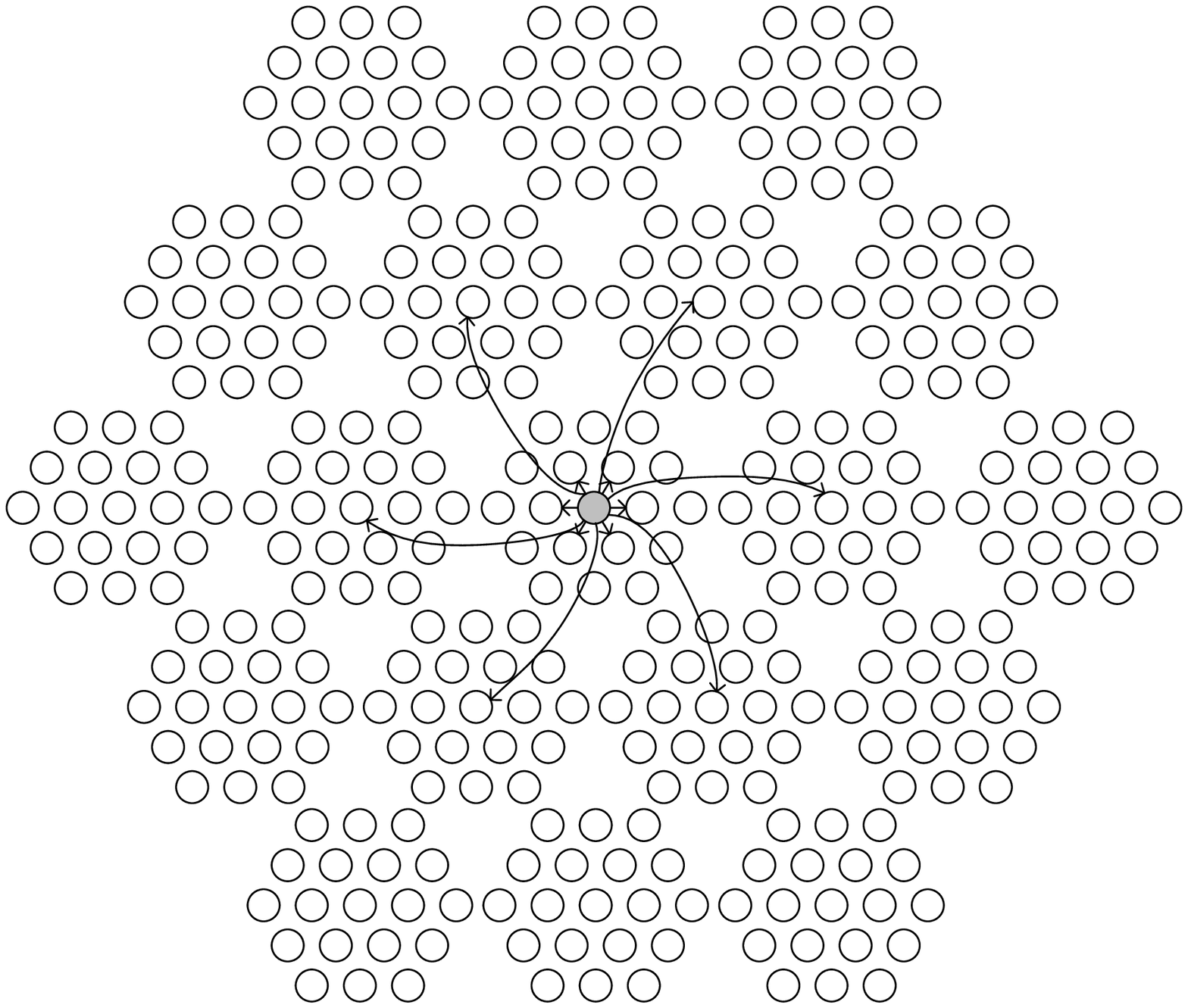}
\caption{Step 1 of the proposed one-to-all broadcast in $EJ_{2+3\rho}^{(2)}$.}
\label{2+3rho_bc_2D_proposed_step1}
\end{figure}

\begin{figure}[!ht]
\centering
\includegraphics[scale=0.45]{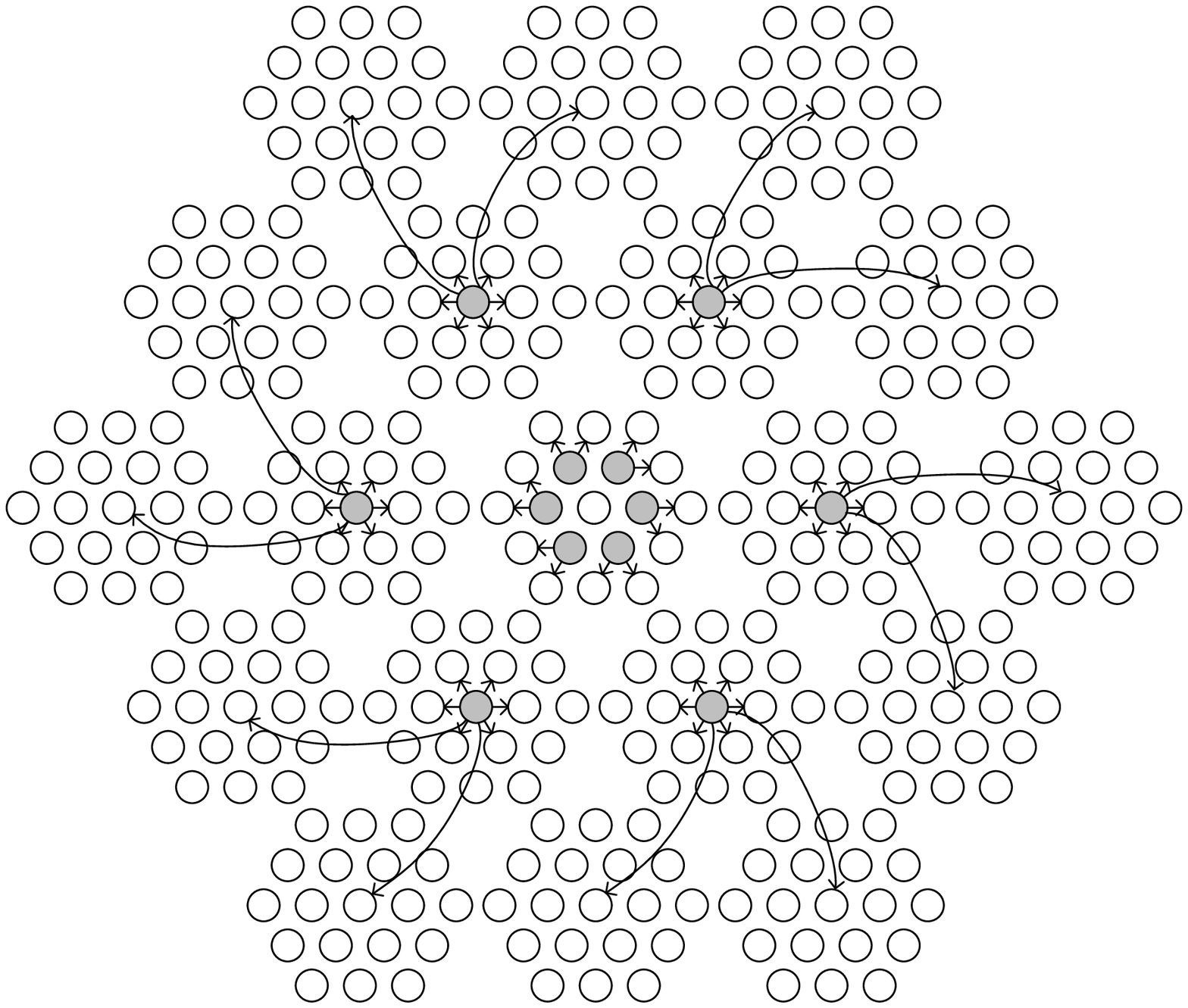}
\caption{Step 2 of the proposed one-to-all broadcast in $EJ_{2+3\rho}^{(2)}$.}
\label{2+3rho_bc_2D_proposed_step2}
\end{figure}

For example, consider the network $EJ^{(2)}_{2+3\rho}$. The improved one-to-all broadcasting works as follows. In step 1, as illustrates in Figure \ref{2+3rho_bc_2D_proposed_step1}, the center node (node 0) sends its message to all of its neighbors in all dimensions. That is, the node 0 sends its message to nodes (0, 1), (0, $\rho$), (0, $\rho^2$), (0, $-1$), (0, $-\rho$), (0, $-\rho^2$), (1, 0), ($\rho$, 0), ($\rho^2$, 0), ($-1$, 0), ($-\rho$, 0), and ($-\rho^2$, 0), i.e., the node 0 calls ONE-TO-ALL($n$,1). After that, in steps $2$ to $nM$, each receiving node is responsible in propagating the message to its sector on the same dimension and to the six sectors in all lower dimensions. It means that the receiving node applies recursively the one-to-all on the lower dimensions while continuing in sending the received message to the neighboring nodes in its sector on the same dimension. The second step of the broadcasting is shown in Figure \ref{2+3rho_bc_2D_proposed_step2}.

\begin{algorithm}
\caption{One-to-All Broadcast}\label{one2all}
\begin{algorithmic}[1]
\Procedure{One-to-All}{$dimension,step$} \\
Let $n$ be the number of dimensions and $M$ be the diameter of the network
\State\parbox[t]{\dimexpr\linewidth-\algorithmicindent}{Send via $+\rho$ packet SECTOR($+\rho$, $+1$, $dimension$, $M-1$, $M-1$, $step+1$) \Comment S1\strut}
\State\parbox[t]{\dimexpr\linewidth-\algorithmicindent}{Send via $+\rho^2$ packet SECTOR($+\rho^2$, $+\rho$, $dimension$, $M-1$, $M-1$, $step+1$) \Comment S2\strut}
\State\parbox[t]{\dimexpr\linewidth-\algorithmicindent}{Send via $-1$ packet SECTOR($-1$, $+\rho^2$, $dimension$, $M-1$, $M-1$, $step+1$) \Comment S3\strut}
\State\parbox[t]{\dimexpr\linewidth-\algorithmicindent}{Send via $-\rho$ packet SECTOR($-\rho^2$, $-1$, $dimension$, $M-1$, $M-1$, $step+1$) \Comment S4\strut}
\State\parbox[t]{\dimexpr\linewidth-\algorithmicindent}{Send via $-\rho^2$ packet SECTOR($-\rho^2$, $-\rho$, $dimension$, $M-1$, $M-1$, $step+1$) \Comment S5\strut}
\State\parbox[t]{\dimexpr\linewidth-\algorithmicindent}{Send via $+1$ packet SECTOR($+1$, $-\rho^2$, $dimension$, $M-1$, $M-1$, $step+1$) \Comment S6\strut}
\If{$dimension > 1$}
\State ONE-TO-ALL($dimension - 1$, $step$)
\EndIf
\EndProcedure
\end{algorithmic}
\end{algorithm}

\begin{algorithm}
\caption{One-to-All Sector}\label{one2allSector}
\begin{algorithmic}[1]
\Procedure{Sector}{$major$, $minor$, $dimension$, $x$, $y$, $step$} \\
Let $n$ be the number of dimensions and $M$ be the diameter of the network
\If{$step > nM$}
\State return
\EndIf
\If{$x > 0$}
\State\parbox[1]{\dimexpr\linewidth-\algorithmicindent}{Send via $minor$ packet SECTOR(\\$major$, $minor$, $dimension$, $x-1$, 0, \\ $step+1$)\strut}
\EndIf
\If{$y > 0$}
\State\parbox[t]{\dimexpr\linewidth-\algorithmicindent}{Send via $major$ packet SECTOR(\\$major$, $minor$, $dimension$, $x-1$, $y-1$, \\ $step+1$)\strut}
\EndIf
\If{$dimension > 1$}
\State ONE-TO-ALL($dimension - 1$, $step$)
\EndIf
\EndProcedure
\end{algorithmic}
\end{algorithm}


\subsection{All-to-All Broadcast}
\label{sectionA2A}
The all-to-all broadcast performs the one-to-all broadcast in every node in the network.
That is, all nodes send their messages to all other nodes in the network.
Note that, all nodes cannot perform the one-to-all broadcast simulatneously since no node in the network can send
and receive at the same time using the same link. Thus, the all-to-all communications is divided into three phases
in which each phase is responsible to propogate the message to two sectors.
Algorithms \ref{all2all} and \ref{all2allSector} describe the all-to-all broadcast.

\begin{algorithm}
\caption{All-to-All Broadcast}\label{all2all}
\begin{algorithmic}[1]
\Procedure{All-to-All}{$dimension,step,phase$} \\
Let $n$ be the number of dimensions and $M$ be the diameter of the network
\If{phase = 1}
\State\parbox[t]{\dimexpr\linewidth-\algorithmicindent}{Send via $+1$ packet SECTOR(\\$+1$, $-\rho^2$, $dimension$, $M-1$, $M-1$, \\ $step+1$, $phase$)\strut}
\State\parbox[t]{\dimexpr\linewidth-\algorithmicindent}{Send via $+\rho$ packet SECTOR(\\$+\rho$, $+1$, $dimension$, $M-1$, $M-1$, \\ $step+1$, $phase$)\strut}
\EndIf
\If{phase = 2}
\State\parbox[t]{\dimexpr\linewidth-\algorithmicindent}{Send via $+\rho^2$ packet SECTOR(\\$+\rho^2$, $+\rho$, $dimension$, $M-1$, $M-1$, \\ $step+1$, $phase$)\strut}
\State\parbox[t]{\dimexpr\linewidth-\algorithmicindent}{Send via $-1$ packet SECTOR(\\$-1$, $+\rho^2$, $dimension$, $M-1$, $M-1$, \\ $step+1$, $phase$)\strut}
\EndIf
\If{phase = 3}
\State\parbox[t]{\dimexpr\linewidth-\algorithmicindent}{Send via $-\rho$ packet SECTOR(\\$-\rho$, $-1$, $dimension$, $M-1$, $M-1$, \\ $step+1$, $phase$)\strut}
\State\parbox[t]{\dimexpr\linewidth-\algorithmicindent}{Send via $-\rho^2$ packet SECTOR(\\$-\rho^2$, $-\rho$, $dimension$, $M-1$, $M-1$, \\ $step+1$, $phase$)\strut}
\EndIf
\If{$dimension > 1$}
\State ALL-TO-ALL($dimension - 1$, $step$, $phase$)
\EndIf
\EndProcedure
\end{algorithmic}
\end{algorithm}

\begin{algorithm}
\caption{All-to-All Sector}\label{all2allSector}
\begin{algorithmic}[1]
\Procedure{Sector}{$major$, $minor$, $dimension$, $x$, $y$, $step$, $phase$} \\
Let $n$ be the number of dimensions and $M$ be the diameter of the network
\If{$step > nM$ and $phase = 3$}
\State return
\ElsIf{$step > nM$ and $phase < 3$}
\State ALL-TO-ALL($n$, 1, $phase$+1)
\Else
\If{$x > 0$}
\State\parbox[t]{\dimexpr\linewidth-\algorithmicindent}{Send via $minor$ packet SECTOR(\\$major$, $minor$, $dimension$, $x-1$, 0, \\ $step+1$,\\ $phase$)}
\EndIf
\If{$y > 0$}
\State\parbox[t]{\dimexpr\linewidth-\algorithmicindent}{Send via $major$ packet SECTOR(\\$major$, $minor$, $dimension$, $x-1$, $y-1$, \\ $step+1$,\\ $phase$)}
\EndIf
\If{$dimension > 1$}
\State ALL-TO-ALL($dimension - 1$, $step$, \\ $phase$)
\EndIf
\EndIf
\EndProcedure
\end{algorithmic}
\end{algorithm}

The all-to-all broadcasting algorithm works as follows. In phase 1, all nodes call ALL-TO-ALL($n$, 1, 1), which means that all nodes perform the one-to-all broadcasting on sectors 6 and 1 in all dimensions. In order to do that, all nodes in the network open three of their ports and utilize them to send the messages to the neighboring nodes over the links ($+\rho$, $+1$, and $-\rho^2$). In addition, all nodes open the other three ports and utilize them to receive the message from their neighboring nodes over the links ($-\rho$, $-1$, and $+\rho^2$). In phase 2, similar to phase 1, but the one-to-all broadcasting is performed on sectors 2 and 3 in all dimensions. Each node opens three ports for sending the messages to their neighbors through the links ($-1$, $+\rho^2$, and $+\rho$), whereas, the other three ports are used to receive the messages from the neighbors through the links ($+1$, $-\rho^2$, and $-\rho$). Finally, in phase 3, the one-to-all broadcasting is applied on sectors 4 and 5 in all dimensions. That is, all nodes open and utilize three of their ports to send the message to their neighbors via the links ($-\rho^2$, $-\rho$, and $-1$); and open and use the other three ports to receive the messages from the neighbors via the links ($+\rho^2$, $+\rho$, and $+1$).

\begin{figure}[H]
\centering
\includegraphics[scale=0.25]{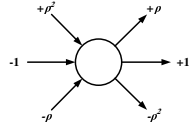}
\caption{The sending and receiving ports in $EJ_\alpha$ during phase 1.}
\label{phase1Ports}
\end{figure}

\begin{figure}[H]
\centering
\includegraphics[scale=0.25]{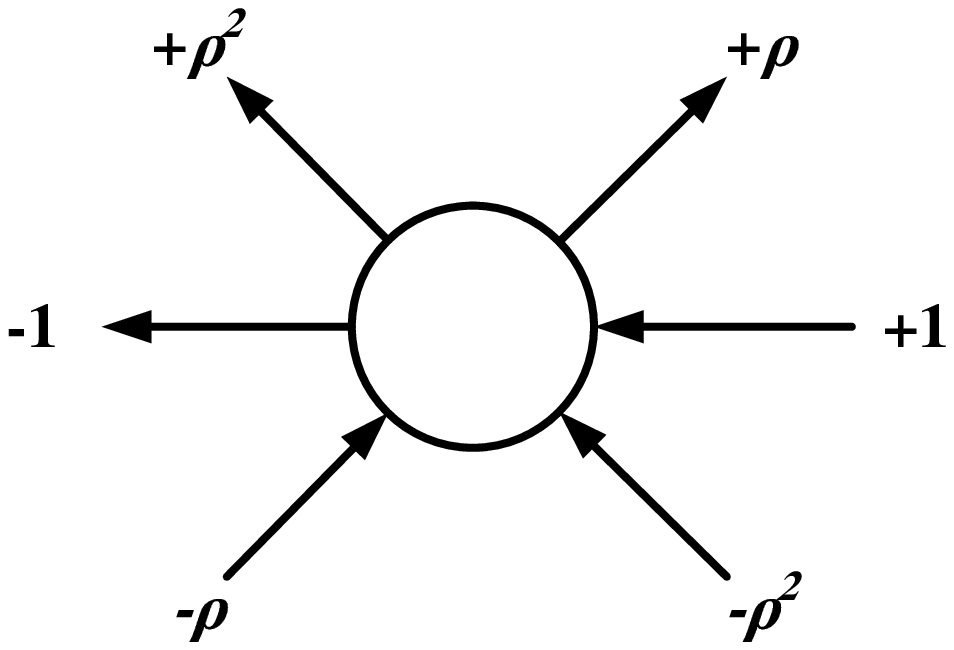}
\caption{The sending and receiving ports in $EJ_\alpha$ during phase 2.}
\label{phase2Ports}
\end{figure}

\begin{figure}[H]
\centering
\includegraphics[scale=0.25]{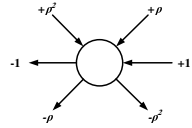}
\caption{The sending and receiving ports in $EJ_\alpha$ during phase 3.}
\label{phase3Ports}
\end{figure}

By the end of phase 3, each node has sent its message to all other nodes in the network. That is, every node in the network has received $N(\alpha)^n-1$ distinct messages. Figures \ref{phase1Ports}, \ref{phase2Ports}, and \ref{phase3Ports} illustrates the ports and the links used by the nodes in a single dimension to send and receive messages in each phase.

\begin{figure}[H]
\centering
\includegraphics[scale=0.75]{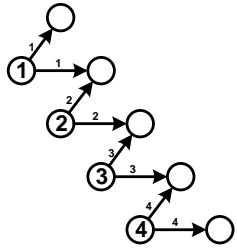}
\caption{All-to-all broadcast: The first step of phase 1 in part of a single dimensional $EJ_\alpha$.}
\label{a2aPhase1Step1}
\end{figure}

\begin{figure}[H]
\centering
\includegraphics[scale=0.75]{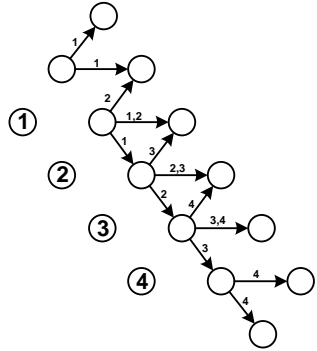}
\caption{All-to-all broadcast: The second step of phase 1 in part of a single dimensional $EJ_\alpha$.}
\label{a2aPhase1Step2}
\end{figure}

\begin{figure}[H]
\centering
\includegraphics[scale=0.75]{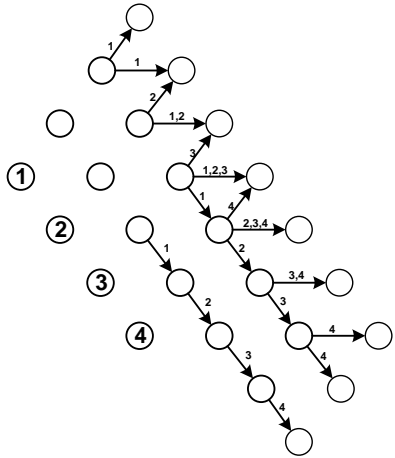}
\caption{All-to-all broadcast: The third step of phase 1 in part of a single dimensional $EJ_\alpha$.}
\label{a2aPhase1Step3}
\end{figure}

In higher dimensional $EJ_\alpha^{(n)}$, the links will be more complex. That is, there will be more links to propagate the messages over the higher and lower dimensions of the network. For clarity, the all-to-all broadcasting in a single dimensional $EJ_\alpha$ is shown in Figures \ref{a2aPhase1Step1}, \ref{a2aPhase1Step2}, and \ref{a2aPhase1Step3} for the first, second, and third steps of phase 1, respectively. Note that, the four senders are numbered with their messages to track the messages in each step and to distinguish between them.

\section{Performance Analysis}
\label{sectionAnalysis}
In this section, the performance of the previous and the proposed one-to-all broadcasting algorithms is analyzed and discussed in term of the number of senders, receivers, active (senders + receivers), and free (total nodes - active nodes) nodes in each step of the broadcasting.

Consider $EJ^{(n)}_\alpha$ where $\alpha \in \mathbb{Z}[\rho]$. Note that, the previous algorithm finishes the broadcast in $nM$ steps where $n$ is the number of dimensions and $M$ is the network diameter (and the number of steps in each round). Then, the number of receiver nodes in each step for the previous broadcasting algorithm is denoted $numR$ and expressed as:
\begin{eqnarray}
numR &=& 6dN(\alpha)^{r-1}
\end{eqnarray}
where $r$ is the round number such that $0 \leq r < n$ and $d$ is the distance or step number during round $r$ such that $1 \leq d \leq M$. Furthermore, the number of sender nodes in each step is equal to the number of receiver nodes of the previous step. That is, the total number of sender nodes in each step is $numS$ and it is computed as:
\begin{eqnarray}
numS &=& 6(d-1)N(\alpha)^{r-1}
\end{eqnarray}
where $r$ is the round number such that $0 \leq r < n$ and $d$ is the distance or step number during round $r$ such that $1 \leq d \leq M$.

Table \ref{simulationIterativeO2a3D} lists the number of free, sending, receiving, and active nodes in each step in the network $EJ^{(3)}_{3+4\rho}$ for the previous one-to-all algorithm. 
\begin{table}
\caption{An analysis of the iterative (previous) One-to-All broadcasting on $EJ^{(3)}_{3+4\rho}$}
\centering
\begin{tabular}{|c|c|c|c|c|c|}
\hline
Round & Step   & Free  & Sending & Receiving & Active \\ \hline\hline
\multirow{3}{*}{1}     & 1      & 50,646 & 1       & 6         & 7      \\ \cline{2-6} 
      & 2      & 50,635 & 6       & 12        & 18     \\ \cline{2-6} 
      & 3      & 50,623 & 12      & 18        & 30     \\ \hline
\multirow{3}{*}{2}     & 1      & 50,394 & 37      & 222       & 259    \\ \cline{2-6}
      & 2      & 49,987 & 222     & 444       & 666    \\ \cline{2-6}
      & 3      & 49,543 & 444     & 666       & 1,110   \\ \hline
\multirow{3}{*}{3}     & 1      & 41,070 & 1,369    & 8,214      & 9,583   \\ \cline{2-6}
      & 2      & 26,011 & 8,214    & 16,428     & 24,642  \\ \cline{2-6}
      & 3      & 9,583  & 16,428   & 24,642     & 41,070  \\ \hline\hline
Total & 12     &       & 26,733   & 50,652     &        \\ \hline
\end{tabular}
\label{simulationIterativeO2a3D}
\end{table}

The following formulas recursively computes the total number of receivers in each step of the proposed one-to-all broadcasting.
\begin{eqnarray}
O2A(n) = \left\{
{\begin{array}{*{20}{l}}
{6S(n, M-1, M-1)}\\
{O2A(n-1) + }\\
{6S(n, M-1, M-1)}
\end{array}} \right .
\begin{array}{*{20}{l}}
{n = 1}\\
{ } \\
{n > 1}
\end{array}
\end{eqnarray}
\begin{equation}
S(n, x, y) = \left\{
{\begin{array}{*{20}{l}}
{S(n, x-1, 0) +} \\
{ S(n, x-1, y-1)}\\
{}\\
{S(n, x-1, 0)}\\
{}\\
{O2A(n-1) + }\\
{ S(n, x-1, 0) + }\\
{ S(n, x-1, y-1)}\\
{}\\
{O2A(n-1) + }\\
{ S(n, x-1, 0)}\\
{}\\
{O2A(n-1)}
\end{array}} \right .
\begin{array}{*{20}{l}}
{}\\
{n = 1, x > 0, y > 0}\\
{}\\
{n = 1, x > 0, y = 0}\\
{}\\
{}\\
{}\\
{n > 1, x > 0, y > 0}\\
{}\\
{}\\
{n > 1, x > 0, y = 0}\\
{}\\
{n > 1, x = 0, y = 0}
\end{array}
\end{equation}
where $n$ is the number of dimensions, $M$ is the network diameter, and $x,y$ are steps counters. The number of senders in the $i^{th}$ step of the proposed algorithm is as follows.
\begin{eqnarray}
\nonumber numS_i &=& numR_{i-1} - number \ of \ S(1,0,0) \\
& & in \ (i-1)^{th} \ step
\end{eqnarray}
where $numS_i$ is the number of the senders in step $i$ and $numR_{i-1}$ is the number of receivers in step $i-1$ calculated based on the the above formulas. Or, $numS_i$ can be computed as follows.
\begin{eqnarray}
numS_i = number \ of \ expanded \ S's \ in \ step \ i-1
\end{eqnarray}

For example, consider $EJ^{(2)}_{2+3\rho}$ where its diameter is $M$ = 2. Then, the step 1 can be written as:
\begin{eqnarray*}
O2A(2) &=& O2A(1) + 6S(2,1,1) \\
       &=& 6S(1,1,1) + 6S(2,1,1)
\end{eqnarray*}
Consequently, the number of receivers is 12, which is the total number of expanded $S$'s in a single step, and the number of the senders is 1. Moreover, each of the $S$ can be expanded in step 2 and it can be written as:
\begin{eqnarray*}
&=& 6(S(1,0,0) + S(1,0,0)) + \\
& & 6(O2A(1) + S(2,0,0) + S(2,0,0))\\
&=& 12S(1,0,0) + 6(6S(1,1,1) + 2S(2,0,0))
\end{eqnarray*}
As a result, the number of receivers is 60 (since there are 60$S$'s) and the number of the senders is 12, which is the number of $S$'s in the previous step excluding $S$(1,0,0)'s. It is possible to get the third step after expanding all $S$'s and removing all $S$(1,0,0)'s from the previous step, i.e., step 2, since the broadcast ends at $S$(1,0,0). This can be expressed as:
\begin{eqnarray*}
&=& 6(6(S(1,0,0) + S(1,0,0)) + 2*O2A(1)) \\
&=& 6(12S(1,0,0) + 12S(1,1,1))
\end{eqnarray*}
Accordingly, the number of receivers is 144 and the number of the senders is 48. Finally, to get the step 4, which is the last step, all $S$(1, 1, 1) are expanded and all $S$(1,0,0) are eliminated since the broadcast finishes at this point. Then,
\begin{eqnarray*}
&=& 6(12(S(1,0,0) + S(1,0,0))) \\
&=& 144S(1,0,0)
\end{eqnarray*}
Thus, the number of receivers is 144 and the number of senders is 72. Since $S$(1,0,0) cannot be expanded then the broadcast ends at this step.

Table \ref{simulationProposedO2a3D} lists the number of free, sending, receiving, and active nodes in each step of the proposed algorithm applied on the network $EJ^{(3)}_{3+4\rho}$. 
\begin{table}
\caption{An analysis of the proposed One-to-All broadcasting on $E{(J)}^3_{3+4\rho}$}
\centering
\begin{tabular}{|c|c|c|c|c|}
\hline
Step    & Free  & Sending & Receiving & Active \\ \hline\hline
1       & 50,634 & 1       & 18        & 19     \\ \hline
2       & 50,491 & 18      & 144       & 162    \\ \hline
3       & 49,807 & 144     & 702       & 846    \\ \hline
4       & 47,593 & 684     & 2,376      & 3,060   \\ \hline
5       & 42,661 & 2,160    & 5,832      & 7,992   \\ \hline
6       & 35,425 & 4,752    & 10,476     & 15,228  \\ \hline
7       & 29,809 & 7,236    & 13,608     & 20,844  \\ \hline
8       & 31,861 & 7,128    & 11,664     & 18,792  \\ \hline
9       & 40,933 & 3,888    & 5,832      & 9,720   \\ \hline\hline
Total   &       & 26,011   & 50,652     &        \\ \hline
\end{tabular}
\label{simulationProposedO2a3D}
\end{table}

\section{Performance Evaluation}
\label{sectionSimulation}
The simulation results of the comparisons between the previous and the proposed one-to-all broadcasting are discussed in this section. The assumption made for the simulation is that the communications are half-duplex and utilize all ports. The number of active nodes, whether sending or receiving, are added up for the purpose of comparisons in each step of the broadcast.

The simulation ran on $EJ^{(n)}_{3+4\rho}$ network such that $2 \leq n \leq 6$. In addition, simulations on $EJ_{1+2\rho}^{(12)}$, $EJ_{2+3\rho}^{(6)}$, $EJ_{3+4\rho}^{(4)}$, $EJ_{4+5\rho}^{(3)}$, and $EJ_{6+7\rho}^{(2)}$ networks are done and taken into consideration since all of these networks have different dimensions but are equal in number of steps. Then, the average of the simulation results is computed. The following is the discussion about these simulations.

\begin{figure}[ht]
\centering
\includegraphics[scale=0.45]{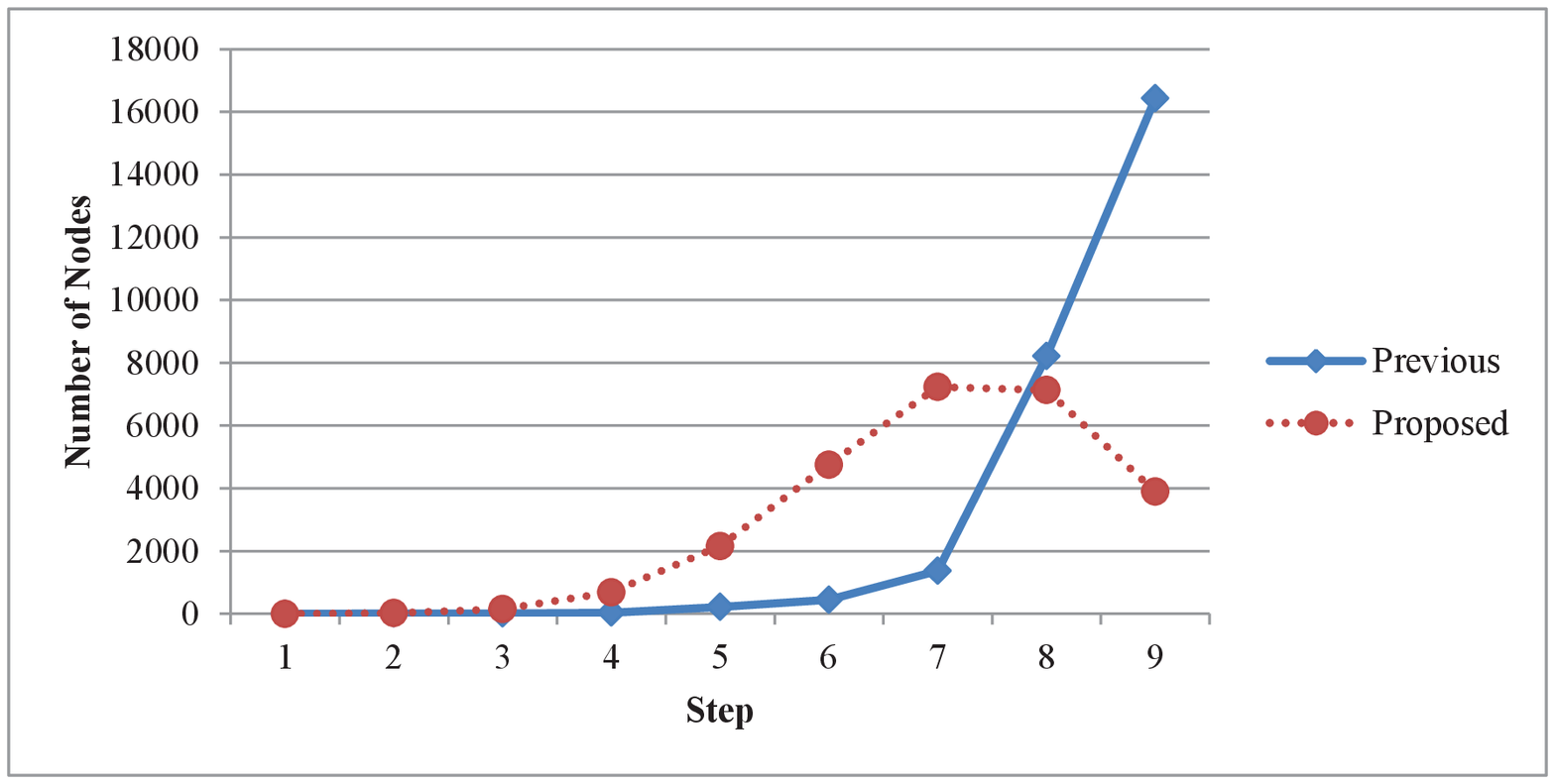}
\caption{One-to-All: Number of senders in each step in $EJ_{3+4\rho}^{(3)}$.}
\label{sendingChart}
\end{figure}

Figure \ref{sendingChart} describes the total number of senders in each step of the broadcasting algorithm in $EJ_{3+4\rho}^{(3)}$. From the figure, the proposed algorithm spreads the message in the middle steps to larger number of nodes than the previous algorithm. Also, it can be seen that the total number of sending nodes in the later steps of the proposed algorithm is less than the previous algorithm. That means, the load on the nodes has been reduced. As a consequence, the proposed algorithm has less overhead than the previous algorithm in later steps.

\begin{figure}[ht]
\centering
\includegraphics[scale=0.45]{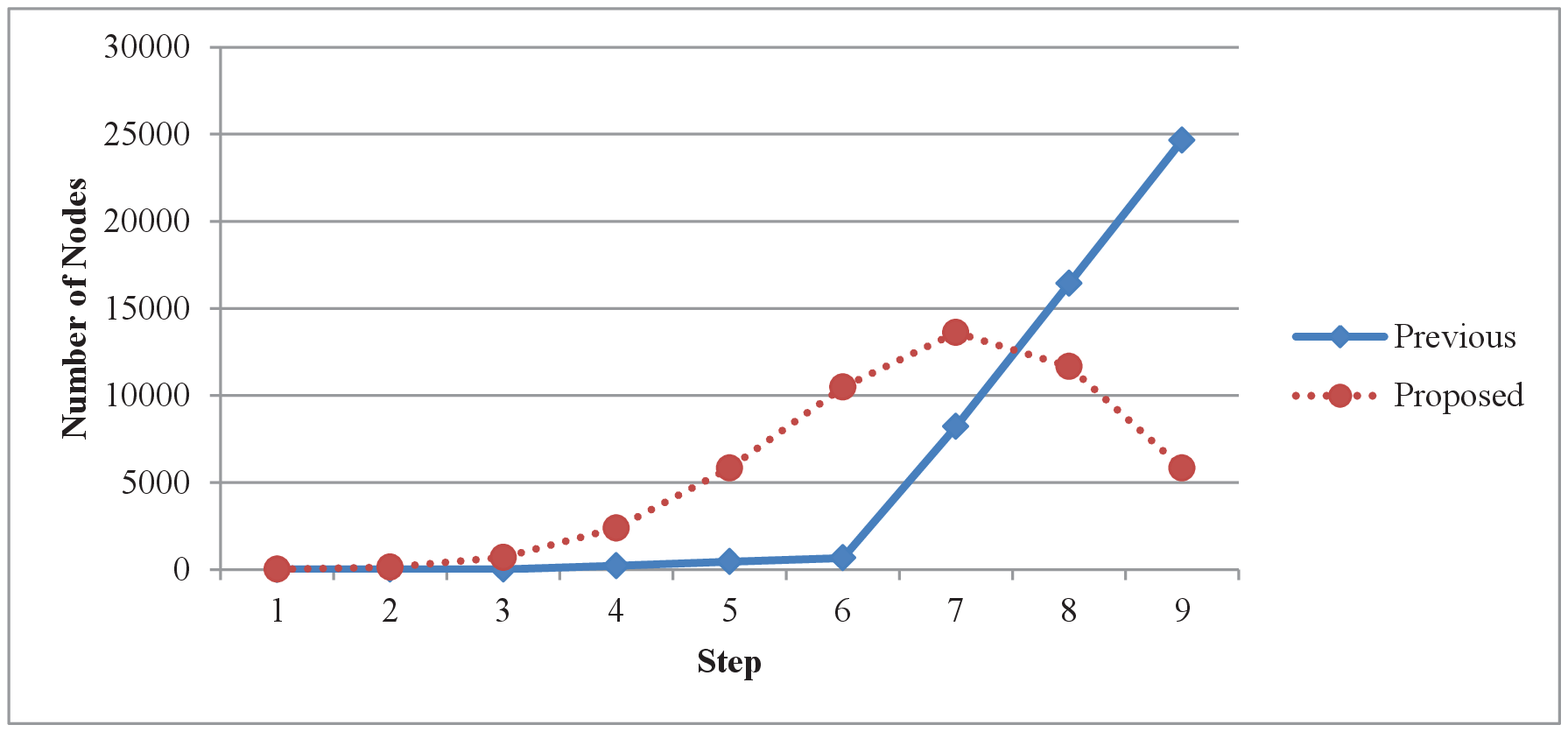}
\caption{One-to-All: Number of receivers in each step in $EJ_{3+4\rho}^{(3)}$.}
\label{receivingChart}
\end{figure}

Figure \ref{receivingChart} shows the number of receivers in every step of the broadcasting algorithm in $EJ_{3+4\rho}^{(3)}$. Compared to the previous algorithm, the proposed algorithm makes most of the nodes had received the message during the middle steps. Thus, the nodes in the middle steps will be available to perform other tasks whether processing, sending, or receiving more messages in the later steps. However, the previous algorithm makes most of the nodes have to wait in order to receive the message during the later steps. Regarding the overhead, it is obvious that most of the nodes are busy in receiving the messages during the later steps of the previous algorithm. As a comparison, the proposed algorithm has less overhead and more free nodes that are available to perform other tasks during the later steps.

\begin{figure}[ht]
\centering
\includegraphics[scale=0.45]{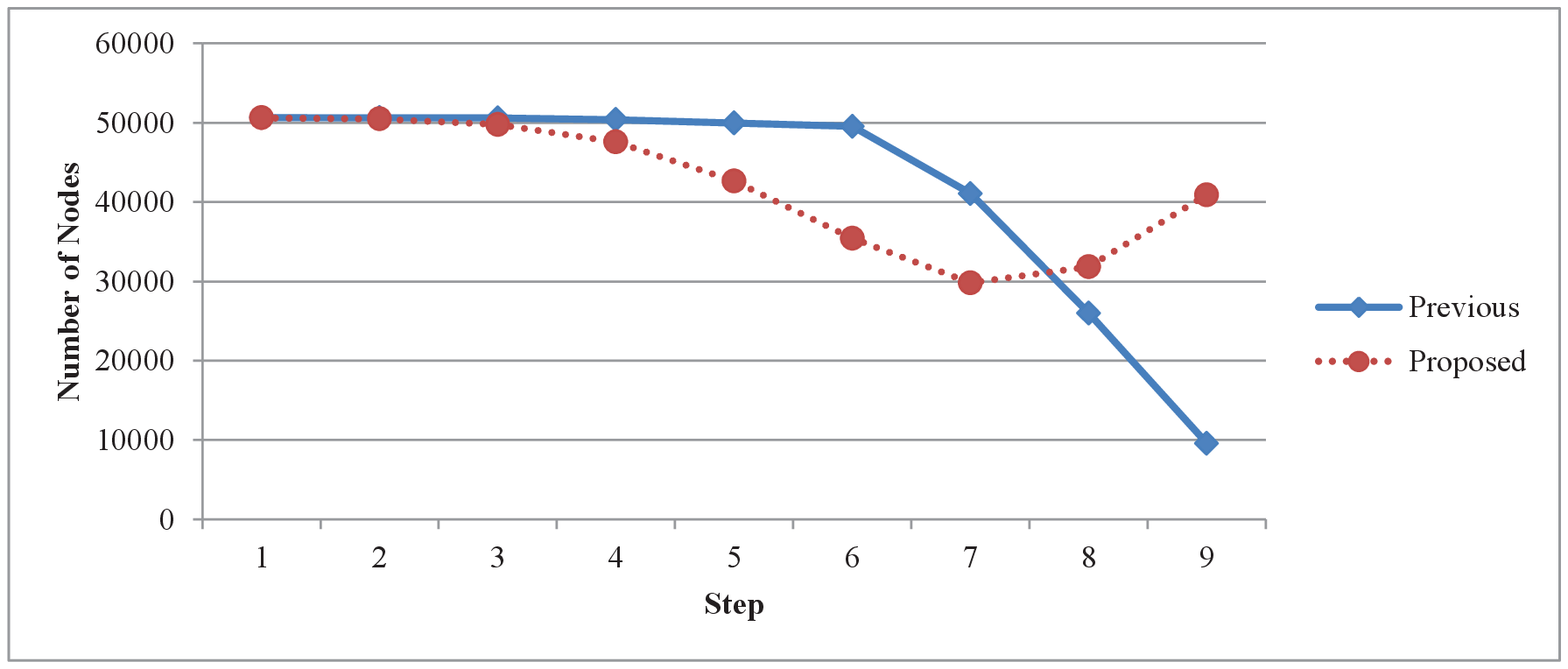}
\caption{One-to-All: Number of free nodes in each step in $EJ_{3+4\rho}^{(3)}$.}
\label{freeChart}
\end{figure}

Figure \ref{freeChart} describes the number of free nodes in each step of the broadcast algorithm in $EJ_{3+4\rho}^{(3)}$. It is clear that the proposed algorithm keeps some nodes busy during the middle steps, which reduces the overhead on the nodes in the later steps.

\begin{figure}[ht]
\centering
\includegraphics[scale=0.45]{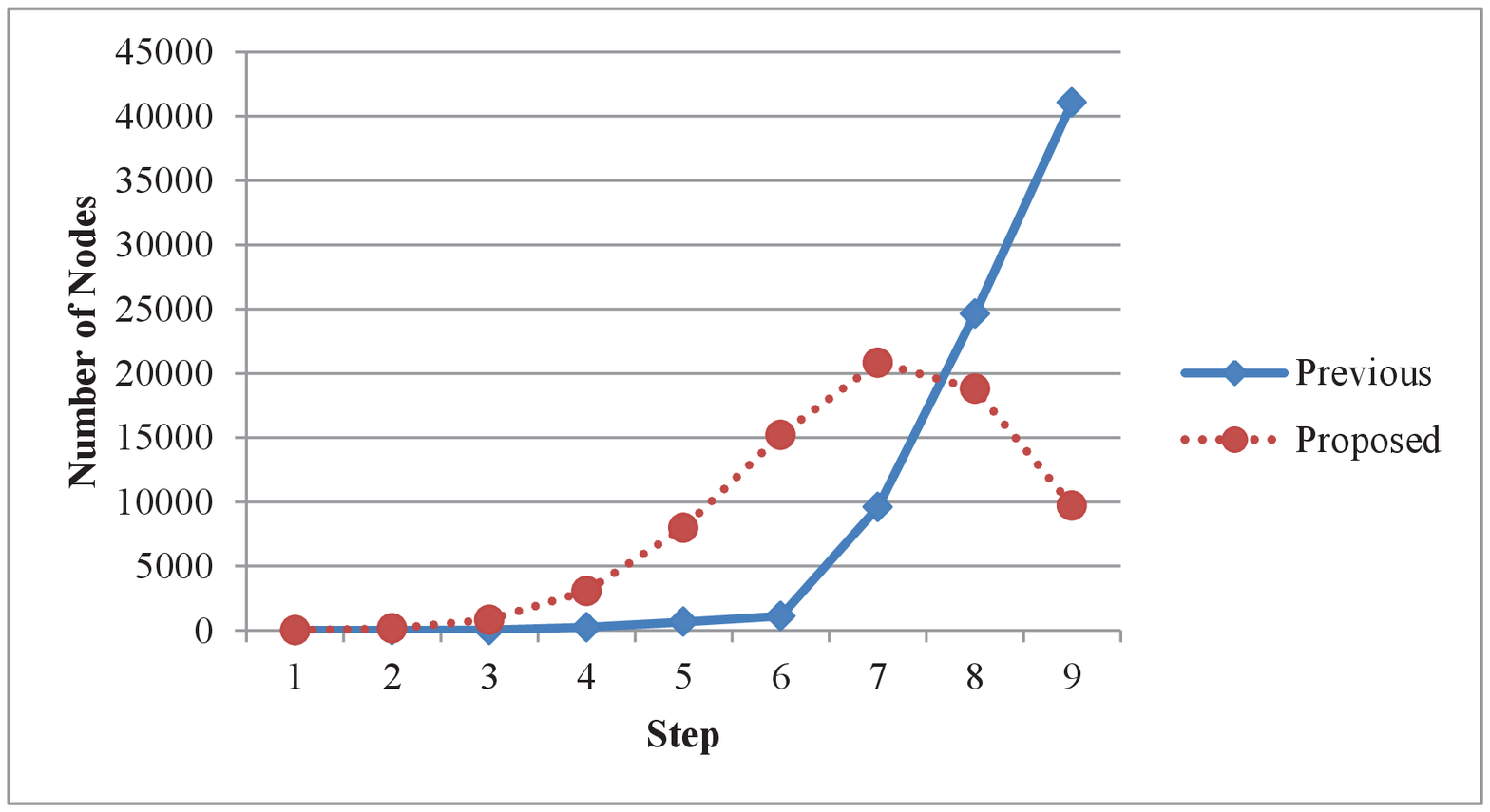}
\caption{One-to-All: Number of active nodes in each step in $EJ_{3+4\rho}^{(3)}$.}
\label{activeChart}
\end{figure}

From Figures \ref{freeChart} and \ref{activeChart}, it can be concluded that the load in the network is distributed between the middle and later steps in the proposed algorithm instead of keeping most of the nodes busy in the later steps.

\begin{figure}[ht]
\centering
\includegraphics[scale=0.30]{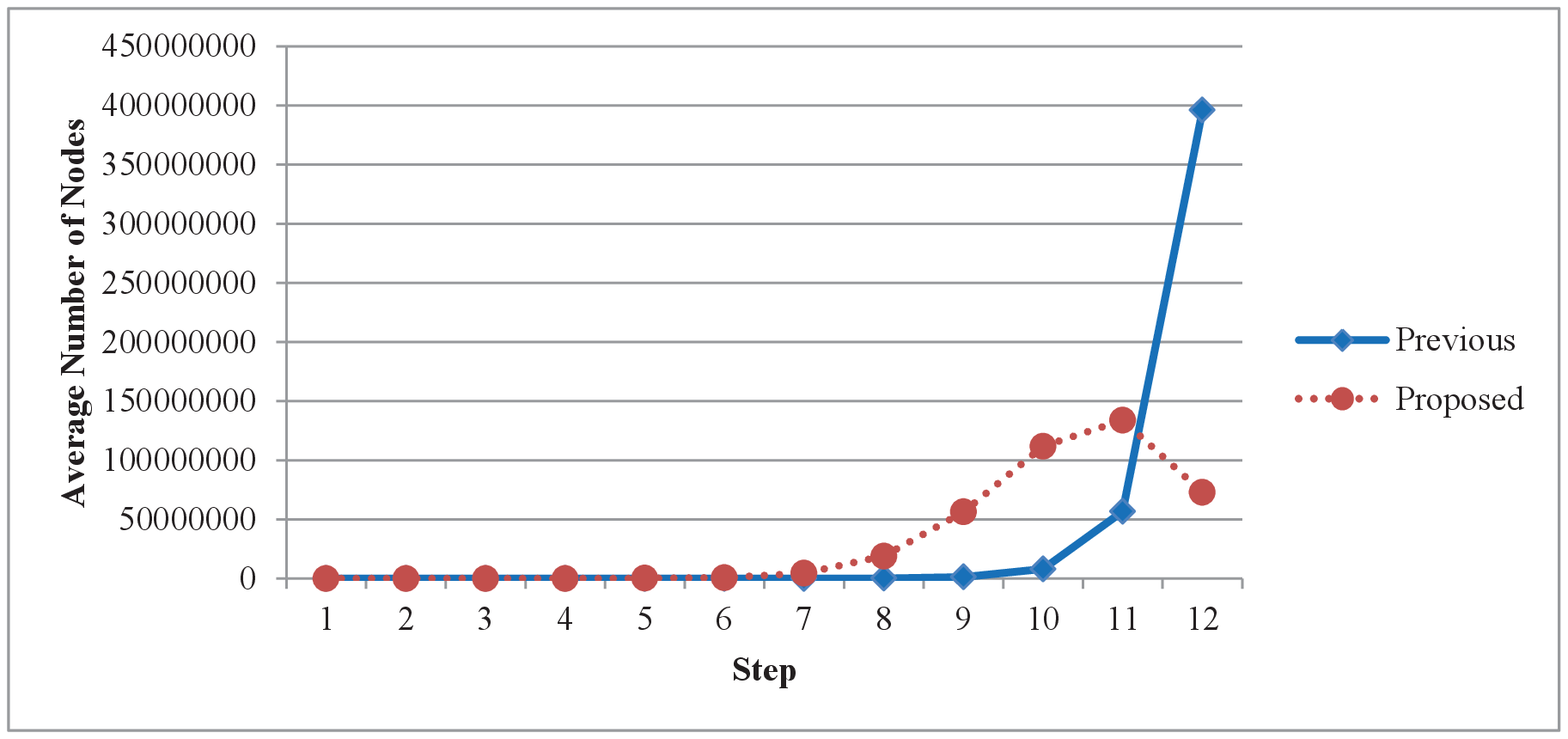}
\caption{One-to-All: Average number of senders in each step.}
\label{avgsendingChart}
\end{figure}

\begin{figure}[ht]
\centering
\includegraphics[scale=0.30]{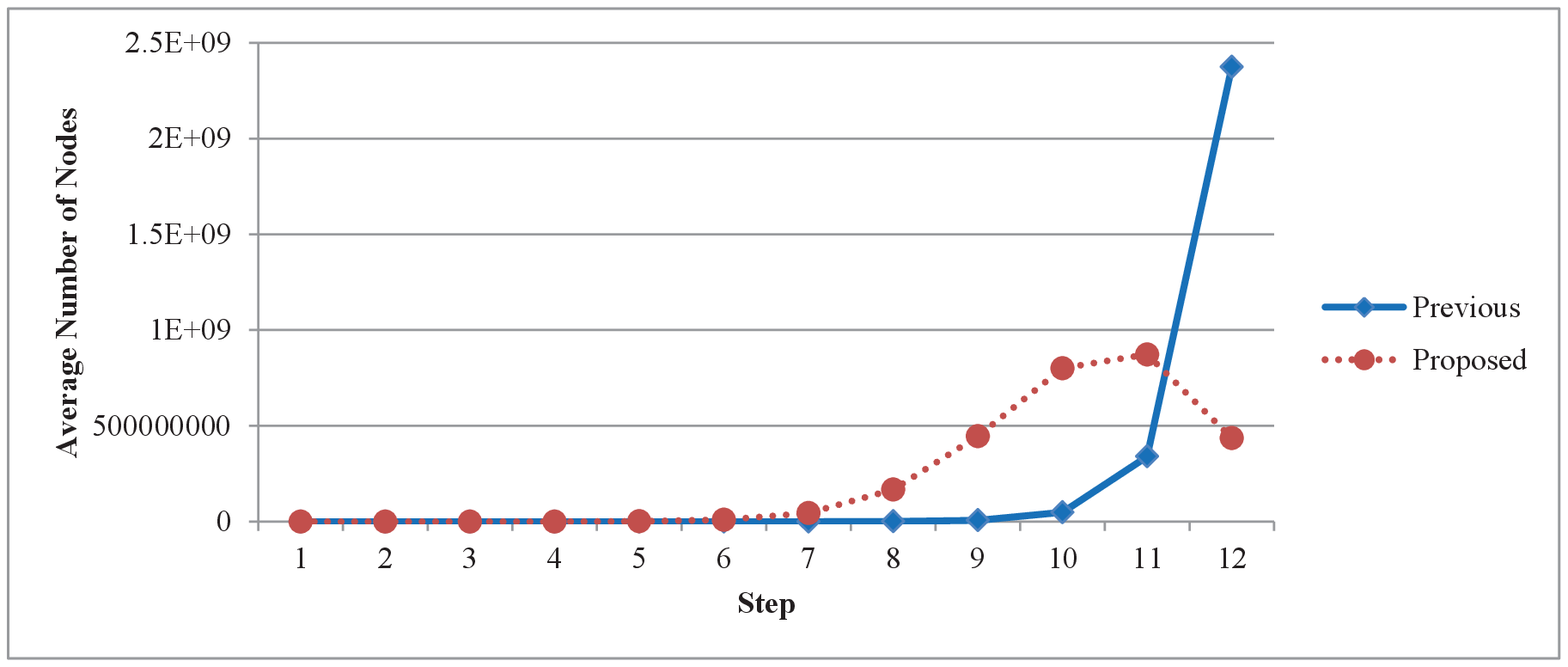}
\caption{One-to-All: Average number of receivers in each step.}
\label{avgreceiverChart}
\end{figure}

\begin{figure}[ht]
\centering
\includegraphics[scale=0.30]{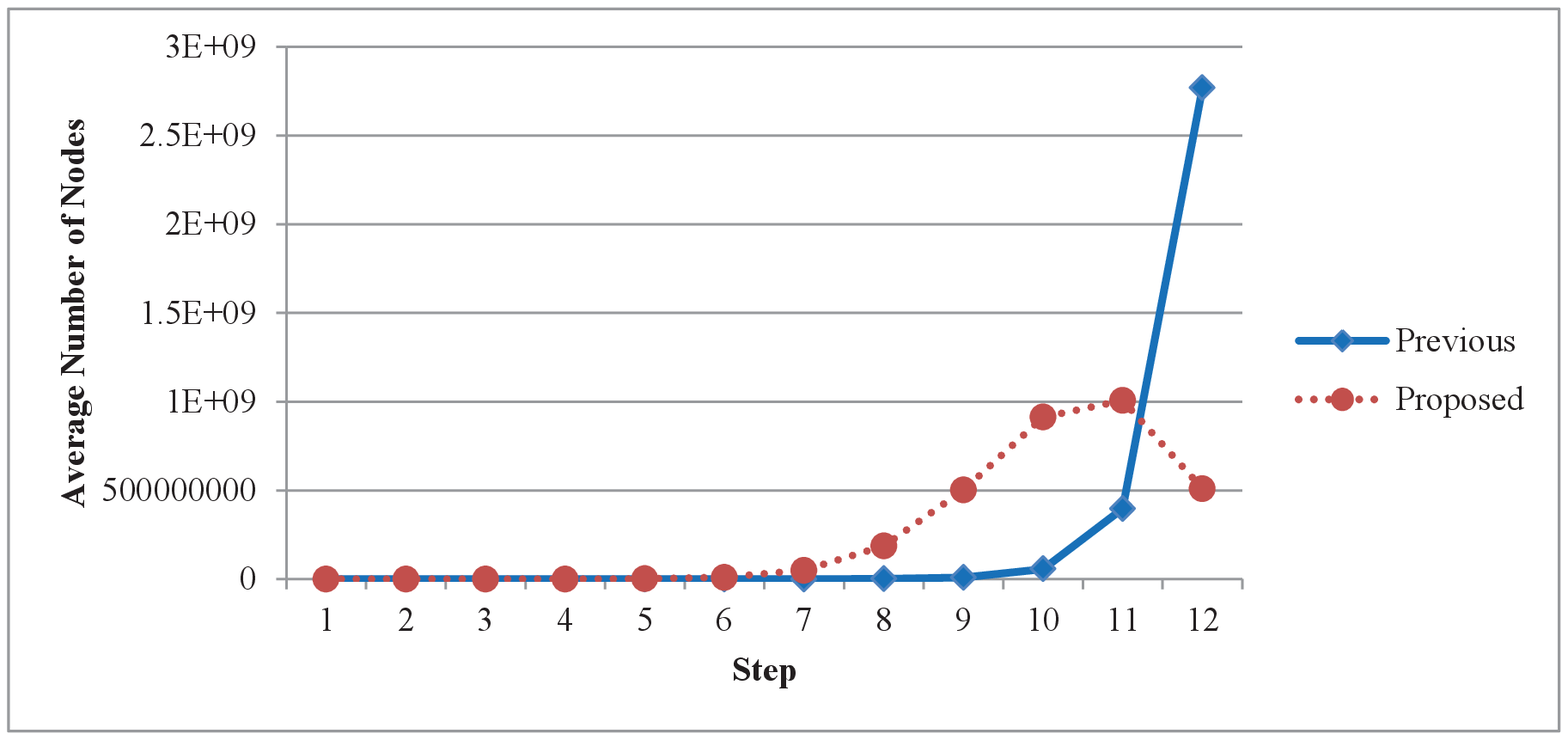}
\caption{One-to-All: Average number of active nodes in each step.}
\label{avgactiveChart}
\end{figure}

The previous and the proposed one-to-all broadcasting algorithms have been ran on the following EJ networks, $EJ_{1+2\rho}^{(12)}$, $EJ_{2+3\rho}^{(6)}$, $EJ_{3+4\rho}^{(4)}$, $EJ_{4+5\rho}^{(3)}$, and $EJ_{6+7\rho}^{(2)}$, since all of them require 12 steps to finish the broadcast. In every step, the average number of sending, receiving, and active nodes have been computed. Figures \ref{avgsendingChart}, \ref{avgreceiverChart}, and \ref{avgactiveChart} illustrate, in respective order, the average number of sending, receiving, and active nodes for both algorithms. Compared with the previous figures, it can been seen that both algorithms have similar plot. As a conclusion, the proposed algorithm is better than the previous one.

\begin{figure}[ht]
\centering
\includegraphics[scale=0.45]{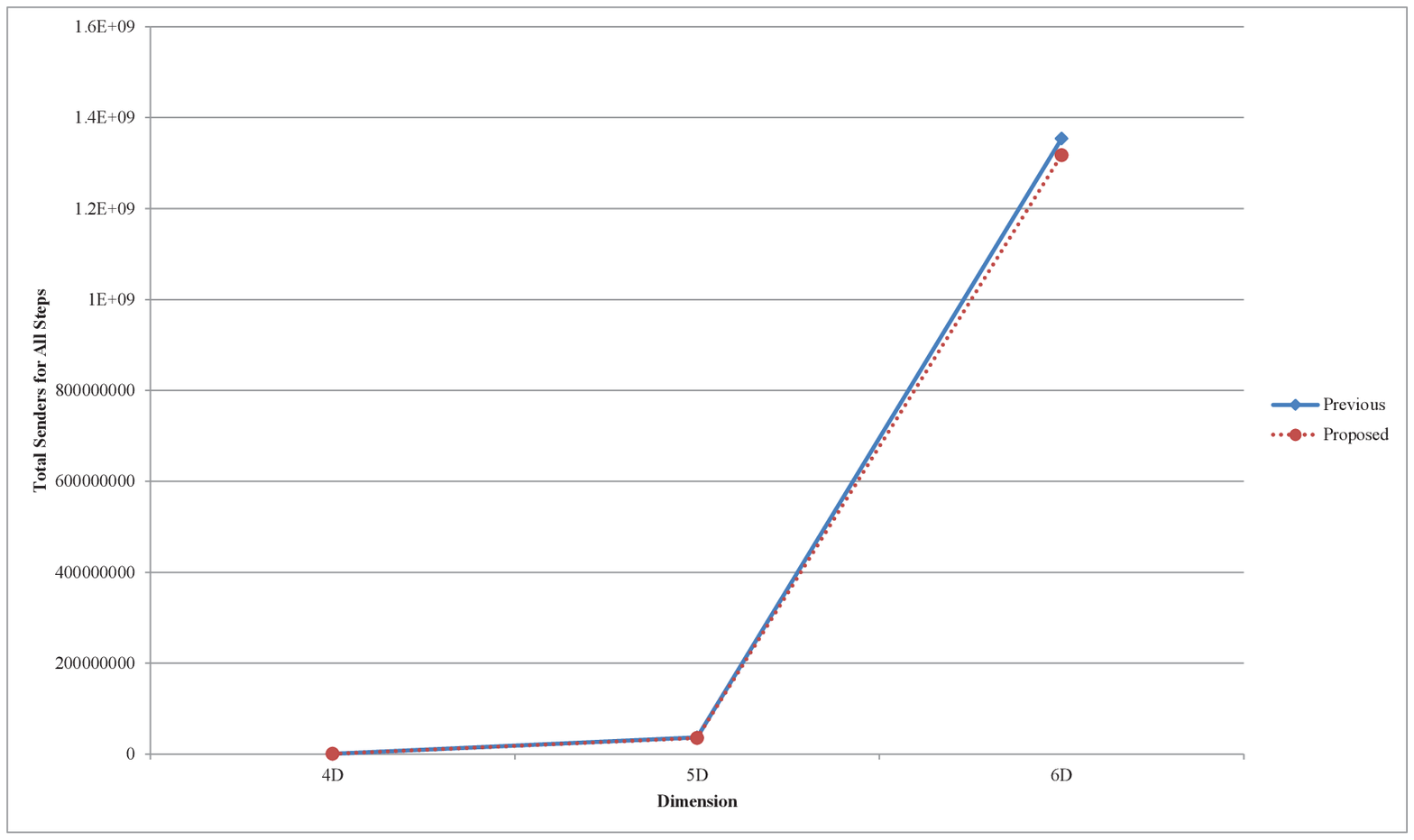}
\caption{One-to-All: The total senders in each step in $EJ_{3+4\rho}^{(n)}$ for $n = 4$ to $6$.}
\label{totalSendersChart}
\end{figure}

In Figure \ref{totalSendersChart}, the proposed one-to-all shows an improvement in the total number of senders in $EJ_{3+4\rho}^{(n)}$ for $n = 4$ to $6$. The difference is that the sender node in the proposed algorithm is used once to send the message to all of its neighbors while it is used several times in the previous one-to-all algorithm in order to send the message to all of its neighbors.

The total number of senders for all steps of the broadcasting in $EJ_{3+4\rho}^{(n)}$ for $n = 1$ to $6$ is listed in Table \ref{simulationSenders} for both, the previous and the proposed, one-to-all algorithms. Further, the table describes the difference between both algorithms in terms of number of senders per dimension. From Figure \ref{totalSendersChart} and Table \ref{simulationSenders}, it can be seen that the proposed one-to-all shows an improvements of approximately 2.7\% for all dimensions. That is, the total number of senders in the proposed algorithm is less than the previous algorithm since in the proposed algorithm the sender is used only once.

\begin{table*}
\caption{Total number of senders in all steps after the broadcast is completed in $EJ^{(n)}_{3+4\rho}$}
\centering
\resizebox{15cm}{!} {
\begin{tabular}{|c|c|c|c|c|c|c|}
\hline
$EJ_{3+4\rho}$	& 1D	& 2D	& 3D	& 4D	& 5D	& 6D \\ \hline\hline
\begin{tabular}[c]{@{}c@{}}Previous\\ One-to-All\end{tabular} & 19 & 722 & 26,733 & 989,140 & 36,598,199 & 1,354,133,382 \\ \hline
\begin{tabular}[c]{@{}c@{}}Proposed\\ One-to-all\end{tabular} & 19 & 703 & 26,011 & 962,407 & 35,609,059 & 1,317,535,183 \\ \hline
Difference & 0  & 19 & 722 & 26,733 & 989,140 & 36,598,199 \\ \hline
Ratio      & 1  & 1.027027027 & 1.027757487 & 1.027777229 & 1.027777763 & 1.027777777 \\ \hline
\end{tabular}
}
\label{simulationSenders}
\end{table*}

\section{Conclusion}
\label{sectionConclusion}
The paper proposes an enhanced one-to-all and all-to-all communication algorithms for higher dimensional Eisenstein-Jacobi (EJ) networks. The proposed one-to-all algorithm is compared with one-to-all algorithm (previous) used in \cite{Hussain2016}. For the comparisons, the author did simulations for both the proposed and the previous algorithms on different sizes of higher dimensional EJ networks and showed that the proposed algorithm achieves 2.7\% less in total number of senders than the previous algorithm. In addition, the simulation results show that proposed algorithm transfer the message to larger number of nodes than the previous algorithm during the middle steps of the broadcasting. As a consequence, the nodes in the middle steps become free and available to process other tasks since the overhead is reduced in the later steps. Further, the broadcasting traffic load in the proposed algorithm is distributed among the steps while it is pushed in the later steps of the previous algorithm.

Furthermore, the paper presented the all-to-all broadcasting algorithm, which is based on the proposed one-to-all algorithm. In all-to-all broadcasting, the communication is assumed to be half-duplex and because of that the all-to-all broadcasting algorithm is divided into three phases where in in each phase the messages are propagated on two sectors.

\nocite{*}

\bibliographystyle{plain}
\bibliography{mybibfile}

\end{document}